\documentclass[reprint,nofootinbib,amsmath,amssymb,prb]{revtex4-2}

\usepackage{physics}
\usepackage{graphicx}
\usepackage{bm}
\usepackage{amssymb}
\usepackage{amsmath}
\usepackage{hyperref}
\usepackage{dsfont}
\usepackage{mathtools}
\usepackage{color}

\allowdisplaybreaks[4]

\newcommand{\AAM}{Aubry-Andr\'{e}}

\begin{document}

\title{Interaction-enhanced many-body localization in a 1D quasiperiodic model with long-range hopping}

\author{Haowei Fan}
\affiliation{Department of Physics, City University of Hong Kong, Kowloon, Hong Kong SAR, China}

\author{Ke Huang}
\affiliation{Department of Physics, City University of Hong Kong, Kowloon, Hong Kong SAR, China}

\author{Xiao Li}
\email{xiao.li@cityu.edu.hk}
\affiliation{Department of Physics, City University of Hong Kong, Kowloon, Hong Kong SAR, China}

\date{\today}

\begin{abstract}
We study the many-body localization (MBL) transition in an 1D exactly solvable system with long-range hopping and quasiperiodic on-site potential introduced in Phys. Rev. Lett. 131, 186303 (2023). 
Unlike other disorder or quasiperiodic model, an interaction-enhanced MBL happens in the moderate interaction regime, which is dubbed as the interaction-enhanced MBL. This counterintuitive phenomenon can be understood by noticing the fragility of the critical band lying at the bottom of the spectrum. The fragile band is localized by other localized states once the interaction is turned on. This mechanism can be verified by introducing a mean-field theory description which can derive highly excited states with high accuracy. The effectiveness of this mean-field theory is captured by the quasihole physics, validated by the particle entanglement spectra.

\end{abstract}

\maketitle


\section{Introduction}
The eigenstate thermalization hypothesis (ETH)~\cite{PhysRevA.43.2046, PhysRevE.50.888} provides a framework for understanding how quantum systems approach thermal equilibrium. However, ETH is known to break down in one-dimensional tight-binding models with random or quasiperiodic disorders, giving rise to many-body localization (MBL)~\cite{doi:10.1146/annurev-conmatphys-031214-014726, RevModPhys.91.021001}. While MBL has been extensively validated through both numerical studies and experiments in finite systems, its behavior in the thermodynamic limit remains an active area of research with many open questions~\cite{PhysRevE.102.062144, PhysRevB.105.174205, PhysRevB.106.L020202, PhysRevLett.130.250405}.

Conventionally, interactions are expected to promote thermalization by enabling energy and information sharing throughout the system. 
This suggests that in interacting systems, stronger disorder is typically required to achieve MBL compared to their noninteracting counterparts~\cite{altman2015universal, PhysRevA.43.2046, RevModPhys.91.021001, PhysRevB.107.035129, PhysRevLett.128.146601}. 
Two prominent examples illustrate this behavior: First, in the famous noninteracting Anderson model~\cite{PhysRev.109.1492}, localization occurs for any finite disorder strength, while the interacting version requires stronger disorder for the MBL phase~\cite{imbrie2016many, PhysRevLett.128.146601}. 
Second, in the noninteracting \AAM\ (AA) model~\cite{AA} with nearest-neighbor (NN) hopping and quasiperiodic potential, the single-particle localization transition occurs at $V_c=1$ [using the convention similar to that in Eq. \ref{sp Hamiltonian} below], whereas the MBL transition is generally believed to occur at a much larger disorder strength~\cite{PhysRevLett.128.146601,PhysRevB.107.035129}. 

A natural question arises: Are there exceptions to this conventional wisdom? In the strong interaction regime, numerical evidence indeed shows enhanced localization~\cite{PhysRevLett.128.146601}, which can be understood through perturbation theory by treating all non-interacting terms as perturbations~\cite{PhysRevLett.128.146601, PhysRevB.107.035129, PhysRevLett.114.100601, PhysRevB.100.214313}. However, the more intriguing regime is that of moderate interactions, where no single term dominates and perturbation theory fails to provide new insight into the problem.

A recent study~\cite{PhysRevResearch.6.L022054} revealed a surprising phenomenon in a generalized \AAM\ (AA) model, also known as the Ganeshan–Pixley–Das Sarma (GPD) model, with single-particle mobility edge (SPME). 
In this model, moderate interactions can actually enhance many-body localization (MBL), leading to an MBL transition at a lower critical disorder strength compared to the noninteracting case. 
This ``early'' MBL phenomenon, or interaction-enhanced MBL, is counterintuitive since interactions typically promote thermalization and suppress localization. 
The key to understanding this interaction-enhanced MBL lies in the presence of a fragile flat extended band in the single-particle spectrum, which emerges due to the SPME. This band is particularly susceptible to external perturbations due to its narrow width compared to other localized bands. When nearest-neighbor interactions are introduced, the localized bands effectively act as an external disorder potential on this flat band. Consequently, the extended flat band becomes localized by these states, resulting in interaction-enhanced MBL. This mechanism has been rigorously verified using a highly accurate mean-field (MF) theory approach.

In this work, we study such interaction-enhanced MBL transition in another 1D exactly solvable quasiperiodic system with long-range hopping. 
This model was proposed in Ref. \cite{PhysRevLett.131.186303}, and it has a highly-tunable parameter space, encompassing extended, localized, and critical phases, separated by nontrivial mobility edges. 
By tuning the parameters, the fragile flat band can be critical rather than extended phase. 
We aim to understand how the presence of single-particle critical states affects the MBL transition, and verify the validity of the mean-field description introduced in Ref.~\cite{PhysRevResearch.6.L022054} in this model. 
Moreover, we aim to extend the previous work~\cite{PhysRevResearch.6.L022054} by studying the particle entanglement spectra (PES) of the many-body eigenstates, which provides a new perspective on understanding the effectiveness of the MF theory. 
 
The rest of this manuscript is organized as follows. In Sec.~\ref{model}, we briefly introduce the 1D quasiperiodic model we consider, including both its single-particle and interacting cases. In Sec.~\ref{earlyMBL}, we diagnose the interaction-enhanced MBL using two powerful diagnostic tools: the entanglement entropy and the mean gap ratio. In Sec.~\ref{FragileBand}, we demonstrate the fragility of the lowest critical band by introducing an additional weak random disorder. Based on this fragility, in Sec.~\ref{Mean-field description}, we employ a MF description to understand the interaction-enhanced MBL transition, with numerical results showing high accuracy. We explain the effectiveness of the MF theory in Sec.~\ref{Effectiveness of the MF theory} and present our conclusions in Sec.~\ref{conclusion}.

\section{Model}
\label{model}
In this section, we firstly give a brief review of the single-particle properties of the model we are considering. After that, we introduce an additional NN interaction to the system.

\subsection{Single-particle properties}
We start from the 1D single-particle model proposed in Ref.~\cite{PhysRevLett.131.186303}:
\begin{align}
\label{sp Hamiltonian}
    H_{0} =t\sum_{j\neq j'} e^{i\alpha(j-j')}e^{-p\abs{j-j'}}c_j^{\dagger}c_{j'} 
    + \sum_{j}V_jn_j,
\end{align}
where $c_j^{\dagger}(c_j)$ is the creation (annihilation) operator at site $j$. 
The first term describes the long-range hopping, in which the hopping strength between two sites are related to their relative distance, modulated by two parameters $\alpha$ and $p$. 
Moreover, we take $t=1$ as the unit of energy throughout this work. 
The second term describes a quasiperiodic on-site potential, where 
\begin{align}
    V_j = 2V\sum_{l=1}^{+\infty}e^{-ql}{\mathrm{cos}}[l(2\pi \tau j+\phi)] \notag\\
    = 2V\dfrac{e^q\cos(2\pi\tau j+\phi)-1}{e^q\cos(2\pi\tau)-1}. 
\end{align}
The derivation of the second line is given in Appendix~\ref{appA}. 
This potential is controlled by the parameters $V$, $q$, $\tau$, and $\phi$. 
We fix $\tau=\frac{\sqrt{5}-1}{2}$ in this work, which produces a quasiperiodic potential in the lattice. 
Moreover, $\phi\in[0,2\pi)$ is a random phase. 
This model can be reduced to some well-known models such as the AA model and the GPD model \cite{PhysRevLett.114.146601} by tuning the parameters. 
For example, the system reduces to the AA model when both $p$ and $q$ are large. 

\begin{figure}[t]
\centering
\includegraphics[width=0.5\textwidth]{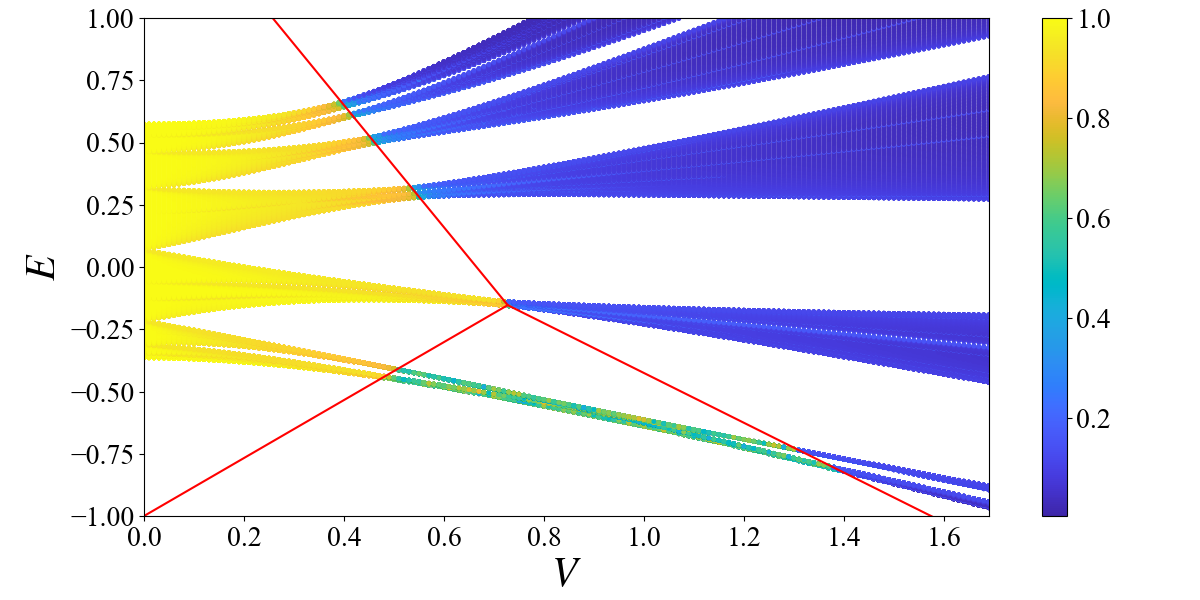}
\caption{The fractal dimension of the model in the single-particle regime. We take $L = 1597$ and utilize rational approximation for the irrational frequency $\tau$ to avoid the localized edge states. The parameters are chosen as $(p, q, \alpha,\phi)=(1.5, 1, 1, 0)$. The red lines defining the SPMEs are given by Eq.~\eqref{SPME}.}
\label{FD_sp}
\end{figure}

One important feature of this model is that it exhibits a rich phase diagram, including extended, localized, and critical phases. 
More importantly, these phases are separated by nontrivial SPMEs.  
Specifically, the exact SPMEs can be analytically derived from the self-duality of the model~\cite{PhysRevLett.131.186303} and given by the following set of equations:
\begin{align}
\label{SPME}
E_\text{EL}&=(V - 1)\frac{\sinh(q)\mp\sinh(p)}{\cosh(q)\mp\cosh(p)} -t,\nonumber\\
E_\text{EC}&=- V\frac{ \sinh(q)\mp\sinh(q)+ 1}{1\mp\cosh(q)} -t,\nonumber\\
E_\text{LC}&= -t\frac{\pm\sinh(p)\mp\cosh(p) +1}{1\mp\cosh(p)} - V, 
\end{align}
which represents the boundaries between extended/localized, extended/critical, and localized/critical states in the energy spectrum, respectively. 

To visualize the different phases and mobility edges in this model, we calculate the fractal dimension $\Gamma$ of the single-particle eigenstates, as shown in Fig.~\ref{FD_sp}. The fractal dimension $\Gamma_m=-\frac{{\mathrm{ln}}\mathrm{IPR}(m)}{{\mathrm{ln}}L}$ of the $m$-th eigenstate characterizes how the wavefunction is distributed in real space, where the inverse participation ratio (IPR) is defined as ${\mathrm{IPR}}(m)=\sum_j \abs{\psi_{m,j}}^4$. Extended states have $\Gamma\rightarrow1$ while localized states have $\Gamma\rightarrow0$.
With parameters set as $(p, q, \alpha,\phi)=(1.5, 1, 1, 0)$ and system size $L=1597$ under periodic boundary conditions, we observe clear phase separations predicted by the SPMEs. At small $V$, the system is fully extended. As $V$ increases, localized states first emerge around $V=0.4$, followed by critical states appearing near $V=0.48$. Beyond $V=0.73$, extended states completely vanish from the spectrum. Finally, at $V_c=1.39$, the system becomes entirely localized.

\subsection{Interacting many-body Hamiltonian}
Having understood the single-particle properties of the model, we now introduce an interaction term to the system to study the many-body localization transition. 
The many-body Hamiltonian is given by 
\begin{align}
    H = H_{0}+Ut_1 \sum_j n_jn_{j+1},
\end{align}
where $U$ is the interacting strength, with $t_1 \equiv te^{i\alpha}e^{-p}$ (the strength of NN hopping). 
This convention of the interaction strength will provide a better benchmark in interaction cases. 
Moreover, we will consider the half-filling case, where the number of particles is equal to half of the system size, $N=L/2$. 
Finally, we take $(p, q, \alpha)=(1.5, 1, 1)$ and adopt the open boundary condition (OBC) unless otherwise specified. 
Under this choice of parameters, $t_1$ is approximately equal to $0.22t$.

\section{The interaction-enhanced MBL transition\label{earlyMBL}}
We now study the MBL transition of the model by two widely-used standard  diagnostics: the entanglement entropy (EE) and the mean gap ratio~\cite{PhysRevLett.128.146601}. 
To be able to access to larger system for less finite-size effect, instead of the von Neumann entropy, we consider the bipartite second R\'{e}nyi entropy $S$ defined by
\begin{align}
    S=-\ln\qty(\tr_L \rho_L^2)
    =-\ln\qty[\tr_L (\tr_R \ketbra{\psi}{\psi})^2)],
\end{align}
where $\tr_{L(R)}$ denotes the partial trace over the left (right) half of the system. 
We normalize the EE of the eigenstates by the Page value $S_T=(L\mathrm{ln}2-1)/2$~\cite{PhysRevLett.71.1291} so that the EE in systems with different system sizes can be compared on an equal footing. 
The mean gap ratio $\expval{r}$ is defined as the average of the gap ratios
\begin{align}
    r_i= \mathrm{min}\{s_i, {s_i^{-1}}\},\ i\in I,
\end{align}
over the considered spectrum $I$, where $s_i=(E_{i+1}-E_i)/(E_{i}-E_{i-1})$ is the ratio of the two adjacent energy gaps above and below the $i$-th level.

These two quantities are effective diagnostics to distinguish between thermal and MBL phases. In the thermal phase, eigenstates exhibit volume-law entanglement with $S/S_T \approx 1$, and their level statistics follow the GOE with $\expval{r} \approx 0.53$. In contrast, MBL eigenstates show area-law entanglement with $S/S_T \approx 0$ and Poisson level statistics with $\expval{r} \approx 0.386$.

\begin{figure} [t]
\includegraphics[width=0.48\textwidth]{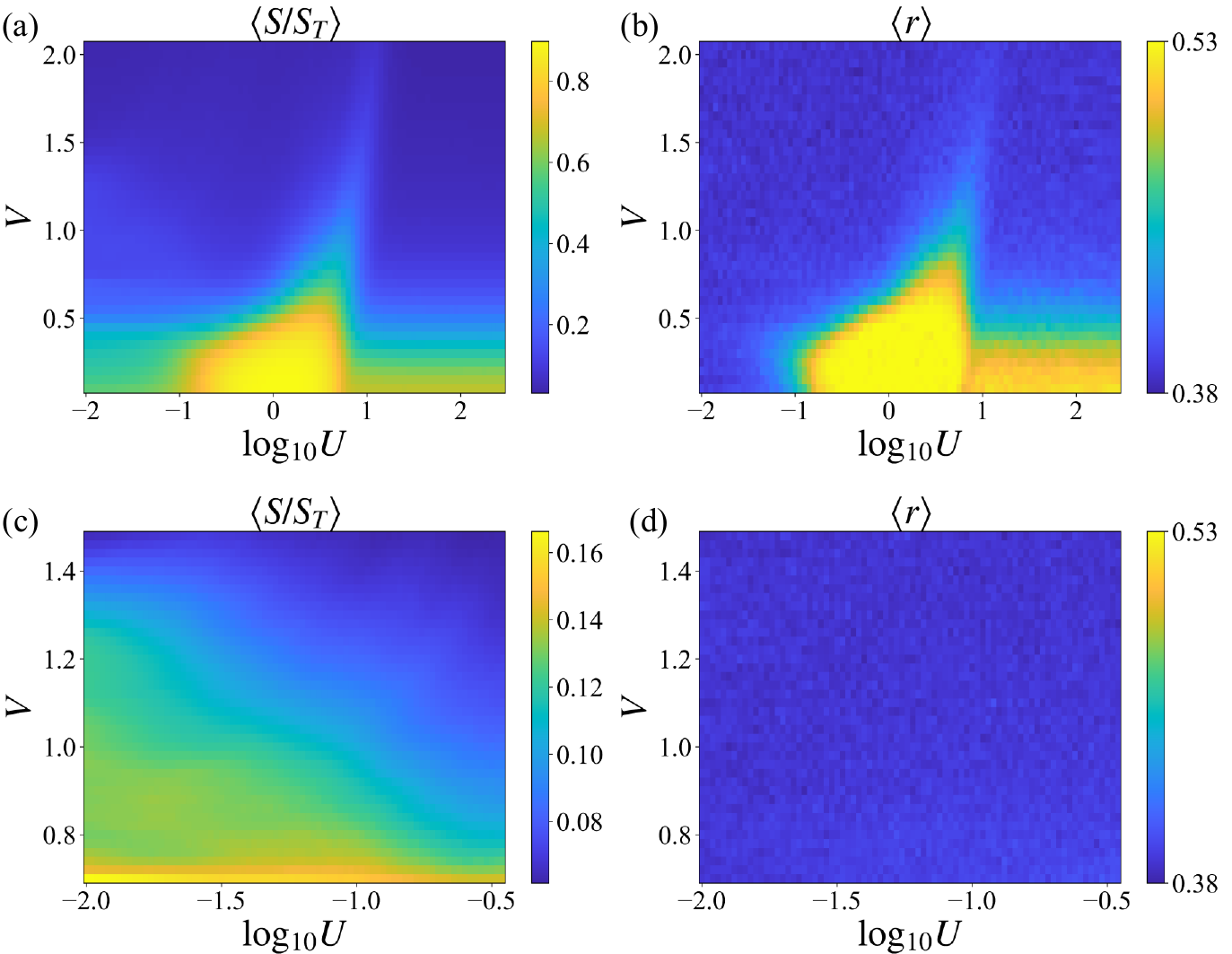}
\caption{
Panel (a) and (b) show the EE and the mean gap ratio as a function of $U$ and $V$ averaged over the whole spectrum. (c) and (d) are the zoomed-in plot of the regime where the interaction-enhanced MBL happens. 
Here, the results are averaged over six random phase realizations in an $L=16$ system.
}
\label{phasediagram}
\end{figure}

\begin{figure} [t]
\centering
\includegraphics[width=0.48\textwidth]{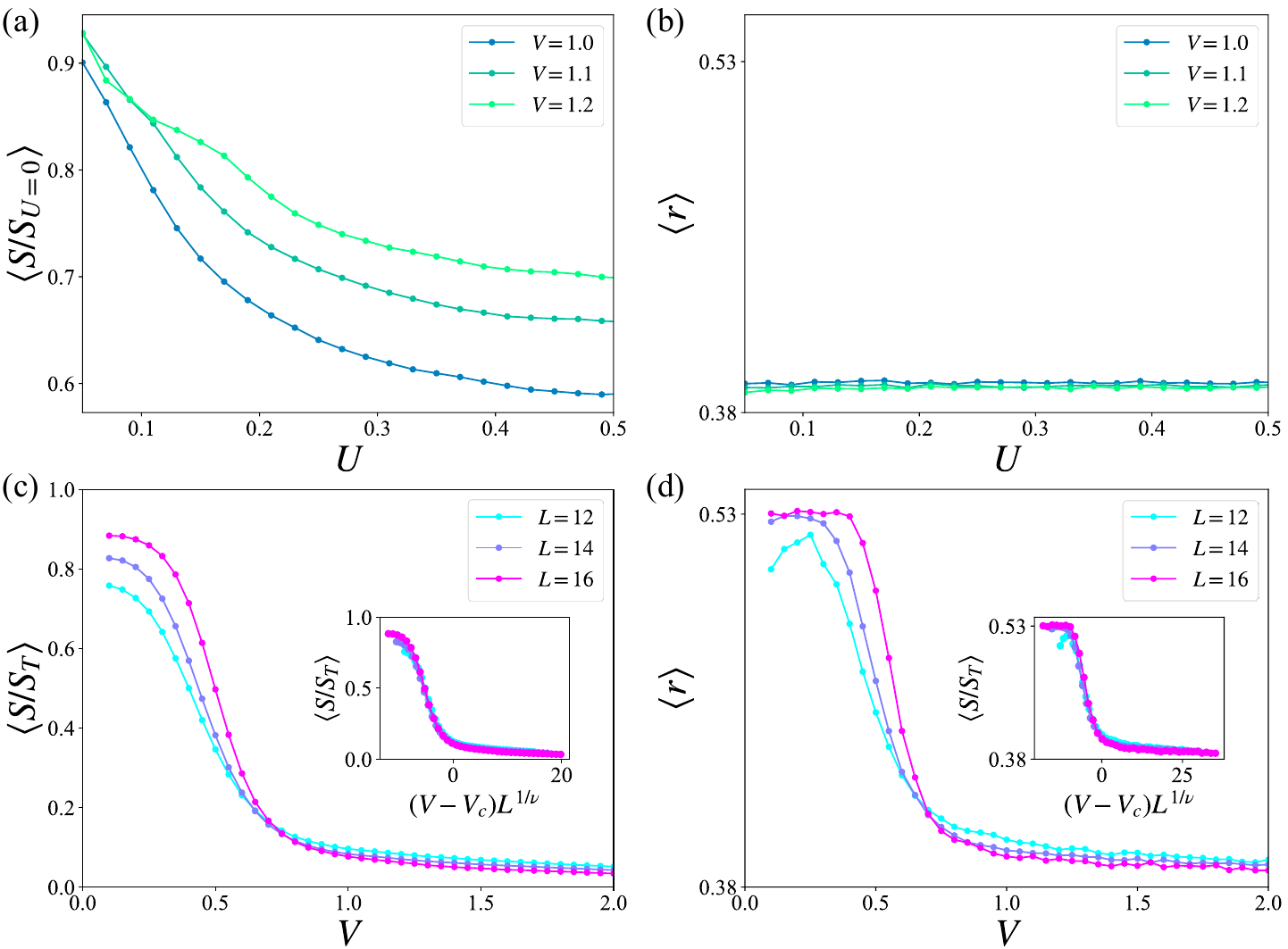}
\caption{
(a) and (b) show the EE and the mean gap ratio averaged over the middle quarter of the many-body spectrum of an $L=16$ system for different $V$. Here the EE are averaged over the middle quarter of the spectrum, and normalized by the numerical noninteracting limit in the finite-size system. 
(c) and (d) show the mean gap ratio averaged over the same spectrum for $U=0.5$ in systems of different sizes. The inset plots show the collapse of data as a function of $(V-V_c)L^{1/\nu}$. We obtain $V_c=0.815$, $\nu=1.009$ for the EE, and $V_c=0.802$, $\nu=0.998$ for the mean gap ratio. The curves are averaged over 10000, 2000, 200 random phase realizations for $L=12$, 14, and 16, respectively.
}
\label{EEr}
\end{figure}

In Fig.~\ref{phasediagram} (a) and (b), we calculate the phase diagrams of the EE and the mean gap ratio as a function of $U$ and $V$ over the whole spectrum in an $L=16$ system. 
In the strong interaction regime $U> \order{10}$, the physics can be viewed as a form of Hilbert space fragmentation~\cite{PhysRevLett.128.146601, PhysRevB.107.035129, PhysRevLett.114.100601, PhysRevB.100.214313}, which is well understood. 
However, something interesting happens in the regime where both critical and localized phases coexist in the noninteracting model ($V \approx 0.73-1.39$). We zoom in this regime in Fig.~\ref{phasediagram}(c) and Fig.~\ref{phasediagram}(d).
The phase diagram shows that the normalized EE decreases as interaction strength $U$ increases from zero to $\mathcal{O}(1)$, while the mean gap ratio remains relatively stable around $0.39$, which is the characteristic of nonergodic phases. 
This behavior provides clear evidence for interaction-enhanced MBL phase in this regime.

To better understand the nature of this interaction-enhanced MBL, we examine the EE and mean gap ratio as functions of interaction strength $U$ for different disorder strengths $V$ in Fig.~\ref{EEr}(a) and Fig.~\ref{EEr}(b). Starting from the noninteracting limit, increasing $U$ leads to a dramatic decrease in EE, indicating enhanced localization. Notably, the mean gap ratio remains consistently around 0.39, characteristic of localized phases, suggesting that the system becomes localized as soon as interactions are turned on.
To study the phase transition more rigorously, we analyze finite-size effects in Fig.~\ref{EEr}(c) and (d) by calculating the EE and mean gap ratio at fixed $U=0.5$ for different system sizes. The data from different system sizes can be collapsed onto universal curves when plotted against the scaling variable $(V-V_c)L^{1/\nu}$~\cite{PhysRevB.89.094516,cond-mat/0505194}. This collapse yields a critical disorder strength $V_c\approx 0.81$, which is significantly lower than the noninteracting localization transition point $V_{c0}=1.39$, providing clear evidence for the interaction-enhanced MBL transition.

Therefore, we have verified the existence of an interaction-enhanced MBL transition in this model which occurs at a critical disorder strength below the counterpart single-particle localization point. 
This phenomenon is quite in contrast with many other well-studied models in which the MBL transition happens ``later'' than that for noninteracting single-particle case \cite{RevModPhys.91.021001, PhysRevLett.128.146601, PhysRevB.107.035129, altman2015universal}. 
In what follows, we will investigate the mechanism behind this interaction-enhanced MBL transition by examining the fragility of the critical band and the effectiveness of the mean-field theory introduced in Ref.~\cite{PhysRevResearch.6.L022054}. 

\section{The role of fragile bands}
\label{FragileBand}
A key mechanism for the interaction-enhanced MBL transition, as demonstrated in Ref.~\cite{PhysRevResearch.6.L022054}, is the presence of a fragile band in the noninteracting system. In our model, the lowest-energy band, which consists of critical states, appears to play this crucial role. Therefore, examining the fragility of this critical band is essential for understanding the interaction-enhanced MBL transition in this model. 

\begin{figure} [t]
\centering
\includegraphics[width=0.5\textwidth]{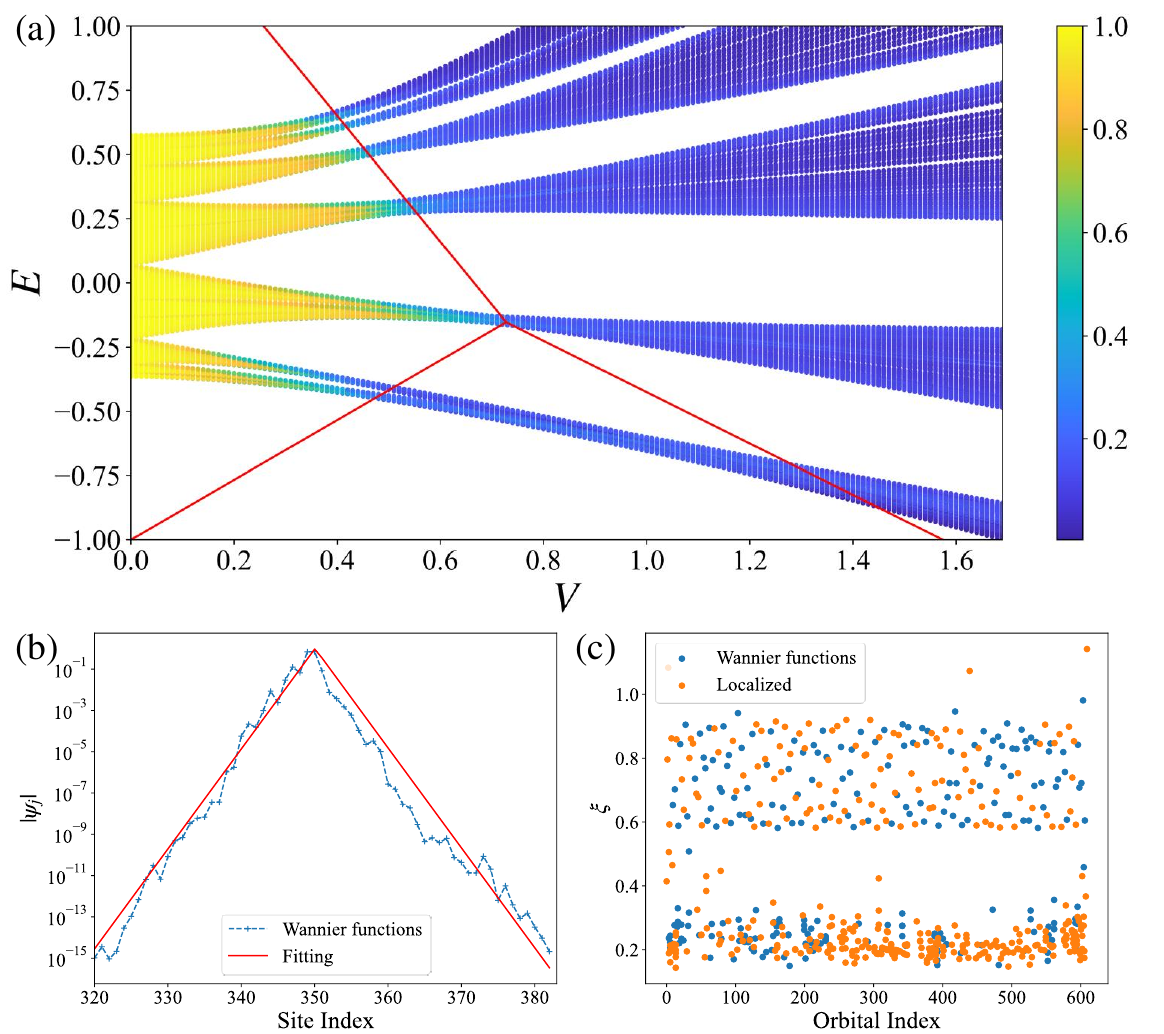}
\caption{(a) The fractal dimension of the single-particle model under the perturbation of Eq. \ref{pert}. (b) The wave function of one Wannier function and its fitting; (c) The localization length of the localized eigenstates and the Wannier functions. In all three panels we take $L=610$ and $\phi=0$. Additionally, in (b) and (c) we take $V=1$.}
\label{FD_pert}
\end{figure}

As discussed in Sec.~\ref{model}, when $V \in (0.73, 1.39)$, most eigenstates are localized except for the lowest-energy band which remains critical. Notably, this critical band has a significantly smaller width compared to the higher-energy localized bands, suggesting its susceptibility to external perturbations. To demonstrate this fragility, we introduce a weak random disorder to the single-particle Hamiltonian:
\begin{align}
\label{pert}
    \delta V = V\sum_j h_j n_j,
\end{align}
where $h_j$ is uniformly distributed in $(-0.02, 0.02)$, approximately $10\%$ of the nearest-neighbor hopping strength $(-t_1, t_1)$. 
As shown in Fig.~\ref{FD_pert}(a), this weak perturbation is sufficient to completely localize the entire spectrum for $V>0.73$.

This localization can be understood through a perturbative analysis.  
Let $\omega$ denote the width of the critical band and $\Delta$ the energy gap between the critical band and other states. When $\omega < \delta V < \Delta$, first-order perturbation theory yields an effective Hamiltonian:
\begin{align}
    H_{\mathrm{eff}} =P\delta VP,
\end{align}
where $P$ projects onto the band of critical states. 
Under this perturbation, the critical states become localized and form a complete basis for the critical band, effectively serving as Wannier functions for this band. 
We quantify the degree of localization by fitting the wavefunctions to the form $\psi_j \propto \mathrm{exp}(-\abs{j-j_0}/\xi)$, where $j_0$ is the localization center and $\xi$ the localization length, as shown in Fig.~\ref{FD_pert}(b). 
As shown in Fig.~\ref{FD_pert}(c), these Wannier functions are deeply localized with $\xi\in(0, 1.2)$. This deep localization in the single-particle regime provides a foundation for the mean-field description we introduce in the following section.

\section{The mean-field description\label{Mean-field description}}
Based on the demonstrated fragility of the critical band, we propose that interactions between particles can induce an effective disorder potential that localizes this band. Specifically, when interactions are turned on, quantum fluctuations from the localized bands act as a disorder potential on the critical band, leading to its localization. To verify this mechanism quantitatively, we employ the MF theory approach developed in Ref.~\cite{PhysRevResearch.6.L022054}.

This MF theory differs fundamentally from conventional MF approaches that typically focus on ground state properties. 
Since MBL is inherently an infinite-temperature phenomenon~\cite{RevModPhys.91.021001}, we need a theory capable of accurately describing highly excited states. The key insight is to use a Slater determinant ansatz that, as we will demonstrate in the following sections, provides remarkably accurate descriptions of the system's excited states. The effectiveness of this approach stems from the deeply localized nature of the single-particle states we identified in the previous section.

\subsection{The Slater determinant ansatz}
The Wannier functions and the other localized orbitals mentioned in Sec. \ref{FragileBand} together form a complete deeply-localized basis. 
In this regime, an argument similar to Ref.~\cite{PhysRevResearch.6.L022054} can be applied. The energy of the orbital and the diagonal part of the interaction, which serves as the effective disorder, dominant the Hamiltonian. The weak tunneling between the Wannier functions and the off-diagonal part of the interaction are perturbative. Therefore, the ansatz that the many-body eigenstates can be regarded as Slater determinants is reasonable, suggesting that the MF theory is effective. 

Specifically, a Slater determinant state $\ket{\Psi}$ is represented by an $L\times N $ matrix $\Psi$ given by
\begin{align}
    \ket{\Psi}\equiv \prod^{N}_{k=1} \bigg[\sum^L_{j=1}\Psi_{jk}c^{\dagger}_j\bigg] \ket{\Omega},
\end{align}
where $\ket{\Omega}$ is the vacuum state, $N$ is the particle number and $\Psi_{jk}$ is the $(j, k)$-entry of an $L\times N$ unitary matrix $\Psi$ satisfying $\Psi^{\dagger}\Psi=\mathds{1}$. 
Each column of $\Psi$ is an occupied orbital. It has been known that a Slater determinant, also known as the Gaussian state, is fully characterized by its one-body reduced density matrix (1-RDM) 
\begin{align}
D_{i, j}\equiv \bra{\Psi}c^{\dagger}_i c_j\ket{\Psi},
\end{align}
which is also called the two-point correlation function of the state \cite{10.21468/SciPostPhys.7.2.024}. In the following discussion, we will use $\Psi$ and $D$ interchangeably, since after direct calculation one obtains that $D=\Psi^*\Psi^{\rm{T}}$.

\subsection{Self-consistent equations}
The MF theory consists of a set of nonlinear self-consistent equations based on the variational principel of Slater determinants. The MF Hamiltonian is given by
\begin{align}
    H_{\rm{MF}}(D) =& H_0+U\sum_j \qty[\bra{\Psi}n_{j-1}\ket{\Psi}+\bra{\Psi}n_{j+1}\ket{\Psi}]n_j \notag\\
    &-U\sum_j \qty[\bra{\Psi}c^{\dagger}_jc_{j+1}\ket{\Psi}c^{\dagger}_{j+1}c_j + \rm{H.c.}]\nonumber\\
    =& t\sum_{j\neq j'} e^{i\alpha(j-j')}e^{-p\abs{j-j'}}c_j^{\dagger}c_{j'} \notag\\
    &+2V\sum_j \sum_{l=1}^{+\infty}e^{-ql}{\mathrm{cos}}[l(2\pi \tau j+\phi)]n_j \notag\\
    &+U\sum_j D_{j,j}(n_{j-1} + n_{j+1}) \notag\\
    &-U\sum_j \qty[D_{j, j+1}c^{\dagger}_{j+1}c_j + \rm{H.c.}],
\end{align}
where the two-body interaction term is factorized into the Hartree and Fock terms by
\begin{align}
    c^{\dagger}_{\nu}c^{\dagger}_{\mu}c_{\mu'}c_{\nu'}
    =& c^{\dagger}_{\nu}c_{\nu'}\langle c^{\dagger}_{\mu}c_{\mu'}\rangle + \langle c^{\dagger}_{\nu}c_{\nu'}\rangle c^{\dagger}_{\mu}c_{\mu'}\nonumber\\
    &-c^{\dagger}_{\nu}c_{\mu'}\langle c^{\dagger}_{\mu}c_{\nu'}\rangle - \langle c^{\dagger}_{\nu}c_{\mu'}\rangle c^{\dagger}_{\mu}c_{\nu'},
\end{align}
and thus is related to the 1-RDM of the Slater determinant. 
The self-consistent equation then reads
\begin{align}
    H_{\rm{MF}}(D)\ket{\Psi} = E\ket{\Psi},
\end{align}
suggesting that each column vector of $\Psi$ is a single-particle eigenstate of $H_{\rm{MF}}(D)$. Therefore, we only need to study the single-particle sector of the $H_{\rm{MF}}(D)$, denoted by $H^{(1)}_{\rm{MF}}(D)$.

The iterative solution of the self-consistent equation begins with an initial product state $\ket{\Psi_0}$. Since we seek highly-excited states rather than ground states, our iteration procedure fundamentally differs from conventional ground-state algorithms. For each iteration step, we calculate $H_{\rm{MF}}(D)$ and select its eigenstate $\ket{\Psi'}$ that maximizes the overlap $\abs{\bra{\Psi}\ket{\Psi'}}$ with the previous iteration's solution. We then update the 1-RDM in $H_{\rm{MF}}(D)$ using this new solution. The iteration continues until the change in overlap between consecutive steps reaches machine precision, at which point we obtain a converged mean-field solution $\ket{\Psi_{\infty}}$ from our initial state $\ket{\Psi_0}$.

\subsection{Accuracy of the MF theory}
\label{Accuracy of the MF theory}
\begin{figure} [t]
\includegraphics[width=0.49\textwidth]{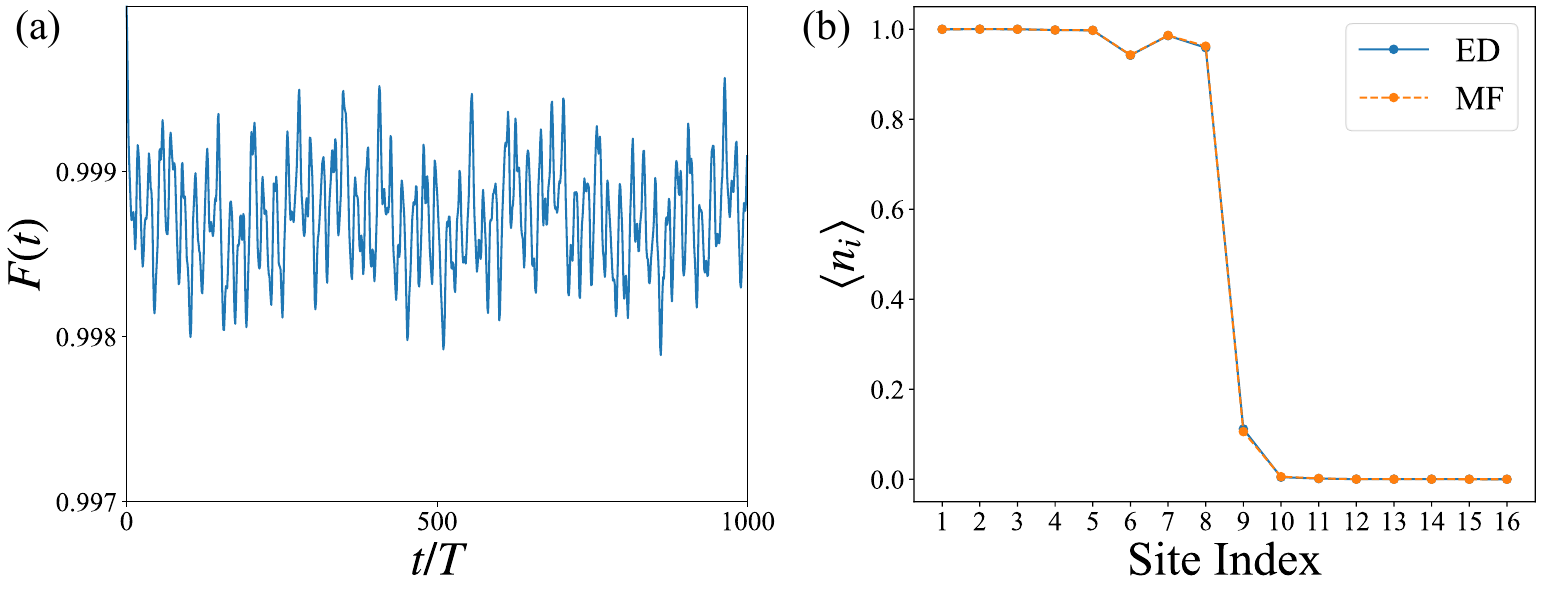}
\caption{(a) shows the fidelity $F(t) = \abs{\bra{\psi}\ket{{\psi(t)}}}$ of the MF-DW state in an $L=16$ system using ED. The evolution step length is one tunneling time. (b) shows the expectation value of the particle number of the MF-DW state and its corresponding many-body eigenstate obtained by ED in an $L=16$ system.}
\label{MF_DW}
\end{figure}

To illustrate the effectiveness of the MF theory, we first focus on a particular MF solution, the MF-domain-wall (DW) state $\ket{\psi_\text{MF}}$ generated by the DW product state $\ket{\psi_0}=\qty[\prod_{j=1}^{N=L/2} c^{\dagger}_{j}] \ket{\Omega}$. To assess the accuracy of this MF solution, we calculate the fidelity $F(t) = \abs{\bra{\psi_\text{MF}}\ket{\psi_\text{MF}(t)}}$ in an $L=16$ system, where $\ket{\psi_\text{MF}(t)}={\rm{exp}}(-iHt/\hbar)\ket{\psi_\text{MF}}$ represents the time evolution under the exact many-body Hamiltonian. 
The fidelity remains remarkably high ($>0.997$) and shows no decay even after $1000$ tunneling times $T=\hbar/t_1$, suggesting that our MF solution closely approximates a true many-body eigenstate. 
To identify this corresponding many-body eigenstate, we use exact diagonalization (ED) to obtain all eigenstates in the $L=16$ system and select the one with maximum overlap ($0.99936$) with our MF solution. 
The nearly identical particle density distributions between this exact eigenstate and the MF state, shown in Fig.~\ref{MF_DW}(b), further confirm the high accuracy of our MF approach.

\section{Effectiveness of the MF theory \label{Effectiveness of the MF theory}}
We note that the MF algorithm we used is modified from the traditional ground-state algorithm, in which one minimizes the energy in each step of iteration to obtain the MF ground state. Generically, highly excited states with close energies may have significant correlation, thus we do not expect the MF algorithm to converge to the desired eigenstate, but rather predict that it will mostly converge to other eigenstates. However, this is not the case as shown in Sec.~\ref{Accuracy of the MF theory} that our algorithm shows high accuracy. Therefore, a natural conjecture arises that the many-body eigenstates that are well predicted by the MF theory are Slater-determinant-like states, so they only allow a few quasihole excitations and, hence, are quite disentangled with their nearby eigenstates. 

The above conjecture can be demonstrated by studying the particle entanglement spectra (PES)~\cite{PhysRevLett.106.100405, PhysRevB.86.165314, Garjani_2015, chandran:hal-00648638}. 
The PES is quite different from the conventional EE computed by cutting the geometry of the system in real space which is to study the correlation across the cutting edge. The PES provides knowledge about the bulk quasihole excitation. For a generic many-body state $\ket{\psi}$, a Schmidt decomposition
\begin{align}
    \ket{\psi} = \sum_{i} e^{-\xi_{i}/2} \ket{\psi^A_{i}}\otimes \ket{\psi^B_{i}}
\end{align}
into particle bipartite region $A$ and $B$ produces the spectrum of $\xi_{i}$, which is called the PES of the state. 
Note that the basis vectors in the two subspaces are required to be normalized:
\begin{align}
    \bra{\psi^A_i}\ket{\psi^A_j}=\bra{\psi^B_i}\ket{\psi^B_j}=\delta_{ij}.
\end{align}
In practice, it is convenient to consider the $n$-body reduced density matrix (n-RDM) 
\begin{align}
    \rho_{\alpha,\beta} = \frac{1}{{N\choose n}} \bra{\psi}c^{\dagger}_{\alpha_1}c^{\dagger}_{\alpha_2}\cdots c^{\dagger}_{\alpha_n}c_{\beta_n}\cdots c_{\beta_2}c_{\beta_1}\ket{\psi}, 
\end{align}
where we divide all particles in the system into two subgroups consisting of $n$ and $N-n$ particles, respectively. Then the PES is obtained by
$\xi_i = -{\rm{log}} \lambda_i$, where $\lambda_i$'s are the spectrum of the singular value decomposition (SVD) of $\rho_{\alpha,\beta}$.

Our MF theory is based on the ansatz that the eigenstates are Slater determinants. As we work in the half-filling subspace, all Slater determinants consist of $N=L/2$ occupied orbitals in a system of size $L$. Quasiholes can only be excited on those occupied orbitals, so we only have $N\choose n$ possible ways for $n$ quasihole excitation. Therefore, we expect that there exists a gap between $N\choose n$-th and $\qty[{N\choose n}+1]$-th levels of the PES since there is no more possible excitation. We take $n=\lfloor N/2 \rfloor$ unless specified otherwise, where  $\lfloor \bullet \rfloor$ stands for the floor function.

\begin{figure} [t]
\centering
\includegraphics[width=0.47\textwidth]{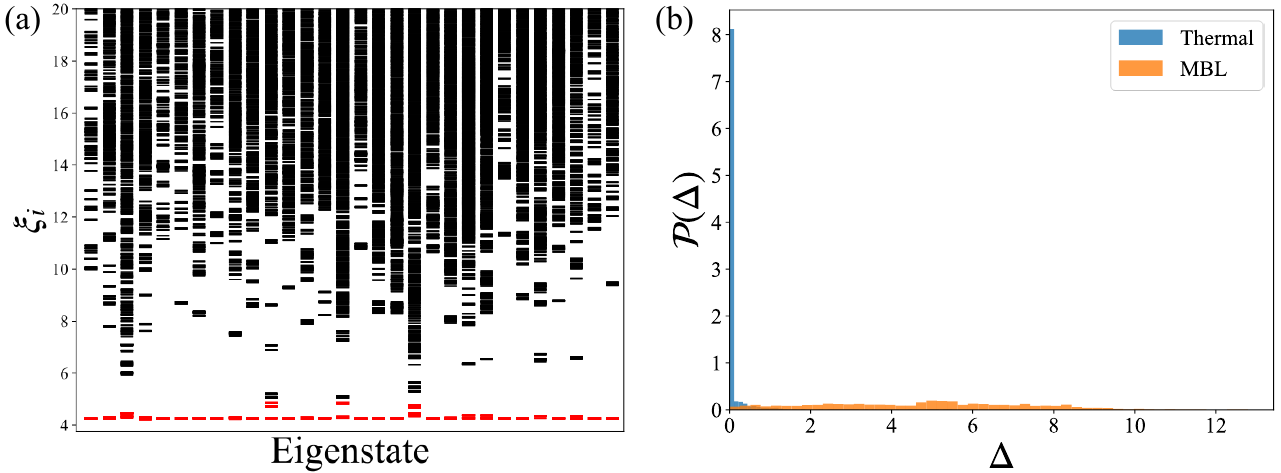}
\caption{(a) shows the PES of 30 eigenstates, which are randomly chosen from the whole spectrum of an $L=16$ system in the MBL phase with $V=1$ and $U=0.5$. We manually mark the levels below the theoretical gap position, i.e., those lowest ${8\choose 4} = 70$ levels, by red color. Note that we omit the levels where $\xi_i$ is greater than 20 because they approach the numerical machine precision. (b) shows the probability density $\mathcal{P}(\Delta)$ of $\Delta$ of the model in both thermal and MBL phases. For the thermal phase, we take $V=0.2$, $U=0.5$. For the MBL phase, we take $V=1$, $U=0.5$. Here, we take the system size as $L=16$.}
\label{PES8406_dis}
\end{figure}

We randomly choose some many-body eigenstates that can be derived from the MF theory with high accuracy, i.e., they have a large overlap with their corresponding MF solutions. We calculate the PES of these eigenstates and find that they are gapped between the predicted two levels as shown in Fig.~\ref{PES8406_dis}(a). To show that this is generically the case, we study the distribution of the value $\Delta = \xi_{{N\choose n}+1} - \xi_{N\choose {n}}$ of all the many-body eigenstates and plot the normalized probability density $\mathcal{P}(\Delta)$ satisfying $\int \mathcal{P}(\Delta) d\Delta =1$ in Fig.~\ref{PES8406_dis}(b), for both thermal and MBL phases. In the thermal phase, most of the $\Delta$ values strongly concentrate around zero, suggesting that most of the states have gapless PES. In contrast, in the MBL phase, most of the states have a finite gap as expected. Due to the wide existence of this gap in the MBL phase, most of the excited eigenstates do not have much entanglement with other eigenstates near them in the energy spectrum, although the energy spectrum is dense. Therefore, the MF algorithm tends to converge to the desired eigenstates due to this property of the energy spectrum.


\section{conclusion}
\label{conclusion}
In conclusion, in this work, we discussed the counterintuitive interaction-enhanced MBL transition in a 1D quasiperiodic model which has highly nontrivial SPME separating the extended, localized and critical phases. 
We demonstrate that the interaction-enhanced MBL happens due to the fragility of the lowest critical band, which can be localized by other localized orbitals that serve as an external disorder to the critical band once the interaction is turned on. This mechanism is verified by an MF theory description with high accuracy. We noticed that this MF theory can produce solutions with surprisingly high accuracy for highly excited states in the middle of the spectrum, which is not the case for the conventional MF algorithm targeting the ground states. 
Therefore, we believe that in the regime we are considering, although the energy spectrum near a highly excited state is very dense, due to the limited number of the quasihole excitation, it is not very entangled with other eigenstates in its neighborhood, so the MF algorithm can converge to the specific eigenstate we are looking for in most of the time.

\begin{acknowledgments}
This work is supported by the Research Grants Council of Hong Kong (Grants No.~CityU~11300421, CityU~11304823, and C7012-21G) and City University of Hong Kong (Project No. 9610428).
K.H. is also supported by the Hong Kong PhD Fellowship Scheme. 
\end{acknowledgments}

\appendix
\section{An equivalent form of the potential}
\label{appA}
The quasiperiodic potential introduced in Eq. \ref{sp Hamiltonian} is an infinite series. Therefore, in order to obtain the numerical results up to machine precision, an appropriate truncation is needed.
However, we can actually obtain exact numerical results, because the infinite series can be analytically calculated as shown below.

The original form of the potential reads 
\begin{align}
\label{original}
    V_j/V=2\sum_{l=1}^{+\infty}e^{-ql}{\mathrm{cos}}[l(2\pi \tau j+\phi)]. 
\end{align}
For a fixed $j$, we have
\begin{align}
    &\sum_{l=1}^{+\infty}e^{-ql}{\mathrm{cos}}[l(2\pi \tau j+\phi)]\nonumber\\
    =&\mathrm{Re}\sum_{l=1}^{\infty} e^{l(-q+i(2\pi \tau j+\phi))}\nonumber\\
    =&\mathrm{Re}\frac{e^{-q+i(2\pi \tau j+\phi)}[1-\underset{l \to \infty}{\lim}e^{l(-q+i(2\pi \tau j+\phi))}]}{1-e^{-q+i(2\pi \tau j+\phi)}}\nonumber\\
    =&\mathrm{Re}\frac{e^{-q+i(2\pi \tau j+\phi)}}{1-e^{-q+i(2\pi \tau j+\phi)}}\nonumber\\
    =&\frac{e^q\mathrm{cos}(2\pi \tau j+\phi)-1}{e^{2q}-2e^{q}\mathrm{cos}(2\pi \tau j+\phi)+1}.
\end{align}
Thus, the original potential is equivalent to 
\begin{equation}
\label{equivalent}
    V_j/V=2 \frac{e^q\mathrm{cos}(2\pi \tau j+\phi)-1}{e^{2q}-2e^{q}\mathrm{cos}(2\pi \tau j+\phi)+1} n_j.
\end{equation}
This form allows for an accurate evaluation of the quasiperiodic potential. 

\section{The level statistics}
\label{LevelStatistics}
In addition to the EE and the mean gap ratio use in main text, the MBL transition can also be characterized by the level statistics, which measures the distribution of 
\begin{align}
    s_n = (E_n - E_{n-1}) / \langle s \rangle, 
\end{align}
where $\langle s \rangle$ is the average level spacing across the entire spectrum. In the thermal phase, the level statistics follows the GOE, with $P(s) = \frac{\pi}{2}s {\mathrm{e}}^{-\pi s^2 /4}$ with the presence of the time-reversal symmetry. In the MBL phase, the Poisson distribution  $P(s) = {\mathrm{e}}^{-s}$ shows up.

\begin{figure}[t]
\centering
\includegraphics[width=0.4\textwidth]{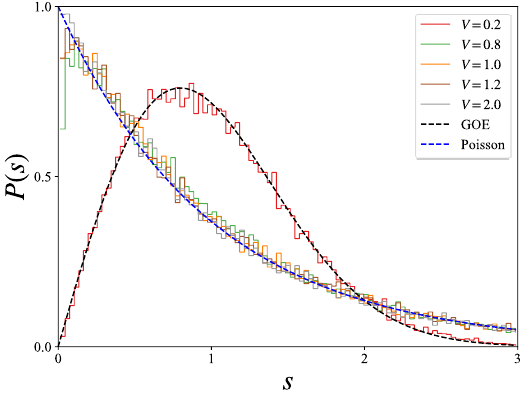}
\caption{The level statistics averaged over the middle half of the spectrum for various $V$ in an $L=16$ system. Here, we take $U=0.5$ and the results are averaged over 10 random phase realizations. The black and blue dash lines represent the GOE $P(s) = \frac{\pi}{2}s {\mathrm{e}}^{-\pi s^2 /4}$ and the Poisson distribution $P(s) = {\mathrm{e}}^{-s}$, respectively.}
\label{LS}
\end{figure}

As shown in Fig. \ref{LS}, for small disorder strength $V=0.2$, the level statistics agree with the GOE distribution, indicating the thermal behavior. For large disorder strength $V>1.0$, the results collapse to the Poisson distribution, which confirms the establishment of MBL phases. For the curve standing for $V=0.8$, the disorder strength is close to the MBL transition point $V_{c}\approx 0.81$, hence it shows intermediate behavior.

\section{Particle entanglement spectrum of some other systems}
\begin{figure}[t]
\centering
\includegraphics[width=0.48\textwidth]{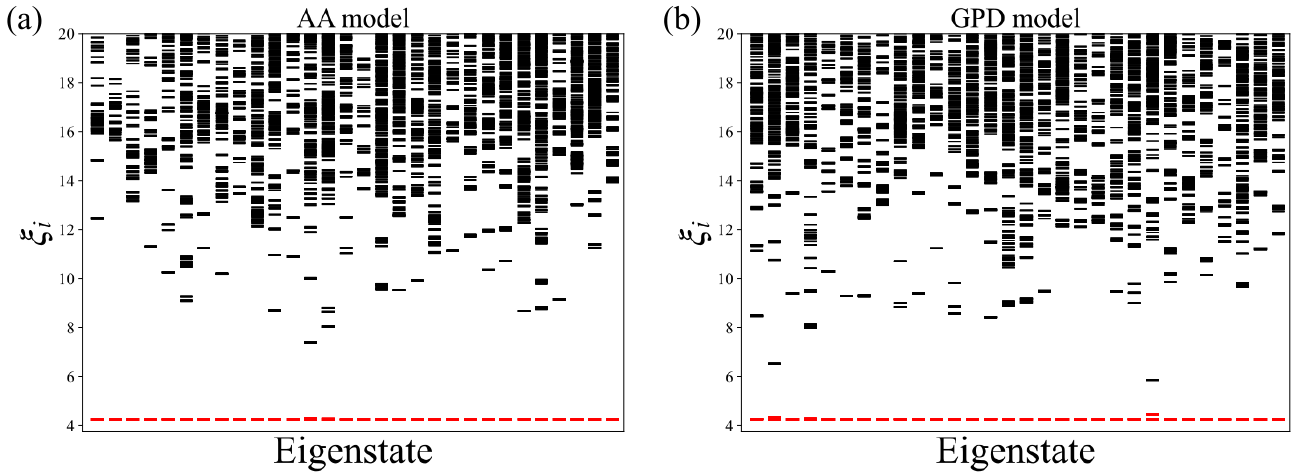}
\caption{The PES of the AA model and the GPD model. In the AA model, we set $U=1$, $V=2.5$, while the MBL transition is known to be localized around $V\approx1.7$. In the GPD model, we set $U=1$, $V=1.5$. This is just the interaction-enhanced MBL phase introduced in~\cite{PhysRevResearch.6.L022054}. In both panel we take the system size $L=16$, randomly choose 30 eigenstates, and mark the levels below the theoretical gap position, i.e., those lowest ${8\choose 4} = 70$ levels, by red color.}
\label{PES_AAandGPD}
\end{figure}

To clarify the effectiveness of our MF description, we also calculate the PES for the AA model and the GPD model in their deeply localized regime in Fig.~\ref{PES_AAandGPD}. Notice that in both cases, the PES is quite gapped in the localized phase, implying that the spectrum is disentangled and the highly excited states can be understood by the quasihole physics discussed in the main text.

\bibliography{ref}

\end{document}